\documentclass[prd,
twocolumn,superscriptaddress,preprintnumbers,nofootinbib]{revtex4}
\usepackage{graphicx}
\usepackage{epsfig}
\usepackage{bm}
\usepackage{latexsym,amssymb,amsmath,amssymb,wasysym,float}

\usepackage{color}

\usepackage{scrextend}
\usepackage{mathrsfs}
\usepackage{enumitem}

\newcommand{\postscript}[2]{\setlength{\epsfxsize}{#2\hsize}
   \centerline{\epsfbox{#1}}}

\usepackage[usenames,dvipsnames]{xcolor}
\definecolor{orange}{cmyk}{0,0.5,1,0}
\definecolor{rossoCP3}{cmyk}{0,.88,.77,.40}
\definecolor{graa}{rgb}{0.8,0.8,0.8}
\definecolor{blaa}{rgb}{0.2,0.2,0.6}

\begin{document}

\preprint{MPP-2025-110}
\preprint{LMU-ASC 14/25}

\title{\color{rossoCP3} Neutrinos from 
  Primordial Black Holes in Theories with Extra Dimensions}
\author{\bf Luis A. Anchordoqui}

\affiliation{Department of Physics and Astronomy,  Lehman College, City University of
  New York, NY 10468, USA
}

\affiliation{Department of Physics,
 Graduate Center,  City University of
  New York,  NY 10016, USA
}

\affiliation{Department of Astrophysics,
 American Museum of Natural History, NY
 10024, USA
}

\author{\bf Francis Halzen}

\affiliation{Department of Physics, Wisconsin IceCube Particle
  Astrophysics Center, University of Wisconsin, Madison, WI 53706, USA}

\author{\bf Dieter\nolinebreak~L\"ust}

\affiliation{Max--Planck--Institut f\"ur Physik,  
 Werner--Heisenberg--Institut,
80805 M\"unchen, Germany
}

\affiliation{Arnold Sommerfeld Center for Theoretical Physics, 
Ludwig-Maximilians-Universit\"at M\"unchen,
80333 M\"unchen, Germany
}

\begin{abstract}
  \noindent The quantum gravity scale within the dark dimension
  scenario ($M_* \sim 10^{9}~{\rm GeV}$) roughly coincides with the energy scale
  of the KM3-230213A neutrino ($E_\nu \sim 10^{8}~{\rm GeV}$). We propose an
  interpretation for this intriguing coincidence in terms of Hawking
  evaporation of five-dimensional (5D) primordial black holes
  (PBHs). 5D PBHs are bigger, colder, and longer-lived than  4D PBHs
  of the same mass. For brane observers, PBHs residing in the higher-dimensional bulk decay
essentially invisibly (only through gravitationally and sterile
coupled modes). As a consequence, constraints on the density of PBHs relative to that of dark matter
from null searches of Hawking evaporation can be avoided. We demonstrate that
Hawking evaporation of 
5D bulk PBHs can explain the KM3-230213A neutrino, 
  evade constraints from upper limits on the gamma-ray flux, and remain
consistent with IceCube  upper limits on the partial decay width of superheavy dark matter particles into neutrinos.
\end{abstract}

\date{May 2025}

\maketitle

\section{Introduction}

Scenarios with large extra dimensions
offer the possibility to describe large hierarchies appearing in particle
 physics and cosmology~\cite{Arkani-Hamed:1998jmv,Antoniadis:1998ig,Montero:2022prj}. Within these scenarios the Standard Model (SM)
fields are confined to a three-dimensional (3D) brane (our universe),
while gravity spills into the bulk of the compact space. In
particular, the dark
dimension scenario provides a framework to
elucidate the origin of the 
cosmological hierarchy problem, because
the anti-de Sitter
distance conjecture in de Sitter
space~\cite{Lust:2019zwm} connects the size of the compact space $R_\perp$ to the dark energy scale $\Lambda^{1/4}$
via $R_\perp \sim \lambda \Lambda^{-1/4}$, where
$\Lambda \sim 10^{-120} M_p^4$ is the cosmological constant, $M_p$ is the
reduced Planck mass, and the proportionality factor is estimated to
be $\lambda \sim
10^{-3}$~\cite{Montero:2022prj}.\footnote{$M_p \simeq 2.38 \times 10^{18}~{\rm
    GeV} \simeq 4.33 \times 10^{-6}~{\rm g}$.}  It is easily seen that if the compact
space has a characteristic length-scale in the micron range, then the
graviton tower of Kaluza-Klein (KK) excitations opens up at
the mass scale $m_{\rm KK} \sim 1/R_\perp \sim {\rm eV}$ and the 5D Planck scale (or species scale
where gravity becomes
strong~\cite{Dvali:2007hz,Dvali:2007wp,vandeHeisteeg:2022btw,Cribiori:2022nke,vandeHeisteeg:2023dlw})
is given by,
\begin{equation}
  M_* \sim m^{1/3}_{\rm KK} \ M_p^{2/3} \sim  10^{9}~{\rm GeV} \, .
\end{equation}

Very recently, the KM3NeT Collaboration  reported the detection of the
highest-energy neutrino ever observed, KM3-230213A~\cite{KM3NeT:2025npi}. The inferred
neutrino energy lies in the range 
\begin{equation}
  10^{7.8} \alt E_\nu/{\rm GeV} \alt 10^{8.9} \, .
\label{KM3Enu}  
\end{equation}
A stunning coincidence is that the energy scale of this neutrino
is just below the 5D Planck scale of the dark dimension scenario. In this paper 
we propose a possible association between these two scales. The proposed
setup provides an explicit example of how black holes could serve as probes for UV physics
closely connected to quantum gravity. This role of black holes
actually emerges in the context of UV-IR mixing within the swampland program for quantum gravity~\cite{Vafa:2005ui}.

Intriguingly, IceCube and the
Pierre Auger Observatory have been
operating with a much larger effective area than KM3Net for a considerably longer
time, yet they have not observed neutrinos with $E_\nu \agt
10^{7.6}~{\rm GeV}$~\cite{IceCube:2025ezc,PierreAuger:2023pjg}. This discrepancy
challenges a cosmogenic neutrino origin for the KM3-230213A 
event. Furthermore, for a diffuse isotropic neutrino flux there is a
$3.5\sigma$ tension between KM3Net and IceCube measurements, which reduces to about
$2.6\sigma$ if the neutrino flux originates in transient
sources~\cite{Li:2025tqf,KM3NeT:2025ccp, Neronov:2025jfj}. 

To understand the origin of KM3-230213A, it is crucial to investigate
consistency with multi-messenger observations, especially 
constraints from gamma-ray experiments. Astrophysical sources of
high-energy neutrinos would also produce a bright gamma-ray
signal. Since there is no convincing evidence of gamma-rays
accompanying KM3-230213A constraints can be placed on a combination of
the source redshift and the intergalactic magnetic field strength
between the source and
Earth~\cite{Fang:2025nzg,Crnogorcevic:2025vou}. In addition, the
absence of a gamma-ray signal challenges scenarios in which
KM3-230213A has a Galactic origin, e.g. through the decay of dark
matter clustered in the halo.

Along this line, a particularly interesting source of KM3-230213A
could be 
the rapid, violent emission of energetic Hawking radiation~\cite{Hawking:1974rv,Hawking:1975vcx} from a
primordial black hole (PBH) near the end of its evaporation
lifetime~\cite{Wu:2024uxa,Klipfel:2025jql}. The Hawking lifetime of a 4D primordial black
hole of initial mass $M$ can be expressed as
\begin{equation}
  \tau_H \sim S \ r_s  \sim \frac{M^3}{M_p^4}  \,,
  \label{4Dtau}
\end{equation}
where $S \sim (M/M_p)^2$
is the entropy of the black hole and $r_s \sim M/M_p^2$ its
Schwarzschild radius. The black hole
temperature is estimated to be~\cite{Hawking:1976de}
\begin{equation}
  T \sim \frac{1}{r_s} \sim M_p \ \frac{M_p}{M} \, .
\label{4DT}
\end{equation}  
Now, Eq.~(\ref{4Dtau}) tells us that only PBHs with an initial mass $M \agt
10^{14.5} {\rm g}$ could survive the current age of the universe, $\tau_{\rm
  universe} \sim 10^{17}~{\rm s}$. In addition, Eq.~(\ref{4DT}) tells
us that black holes of mass $M \sim 10^5~{\rm g}$  would Hawking
radiate at $T \sim 
  10^{8}~{\rm GeV}$, i.e. within the range given in (\ref{KM3Enu}).

The final stage emission is dictated by particle physics~\cite{Halzen:1991uw}. Following~\cite{Klipfel:2025jql} we
assume the degrees of freedom of the particle to be those of the SM, even at very high temperatures. This implies that the
black hole would evaporate at a rate of $\gamma\div\nu = 1\div3$ into photons and neutrinos. Since the PBHs should be clustered in the Galactic halo the
electromagnetic signal cannot be absorbed en route to Earth. Thus, the null
result on photons searches carries in itself a serious challenge for the
interpretation of  KM3-230213A in terms of PBHs evaporating today. Besides,
the energy injection through Hawking evaporation has been used to put
strong constraints on primordial black holes as a dark matter
candidate. An all dark matter interpretation in terms of PBHs is only
possible in the asteroid-mass window: $10^{17} \alt M/{\rm g} \alt
10^{21}$, see e.g.~\cite{Carr:2020xqk,Green:2020jor,Villanueva-Domingo:2021spv}.

Analogous proposals for the origin of
KM3-2313A~\cite{Boccia:2025hpm,Dvali:2025ktz} rely on the assumption
that PBHs experience the memory burden effect~\cite{Dvali:2018xpy,Dvali:2018ytn}, according to which the ``burden'' of memory (essentially
the quantum information stored inside the black hole) resists further
decay by making it energetically unfavorable to offload that
information. The lifetime of a burdened black hole is expected to be
prolonged by the extra powers of the initial entropy
\begin{equation}
  \tau_B \sim S^{1+k} \ r_s \,,
\end{equation}
where $k$ is a positive integer~\cite{Dvali:2020wft}. For $k=1$, the rate of black
evaporation coincides with the total decay width of particle
species~\cite{Basile:2024dqq}. However, the $k=1$ scenario is severely
constrained by observations~\cite{Alexandre:2024nuo,Thoss:2024hsr,Zantedeschi:2024ram,Montefalcone:2025akm,Chianese:2025wrk}. To accommodate the explanation of KM3-230213A one needs to rely on larger values
of $k$, which are more contrived.

A, seemingly different, but in fact closely related subject is 
the inclusion of bulk right-handed neutrinos in theories with large extra
dimensions. The presence of these SM singlets can naturally explain small 4D Dirac masses for the active neutrinos~\cite{Dienes:1998sb, Arkani-Hamed:1998wuz,
  Dvali:1999cn}. Besides, adding a sterile neutrino generation in the
correct mass range to the SM could help ameliorate anomalies of
neutrino oscillation experiments that
cannot be explained by the standard three neutrino
picture~\cite{Diaz:2019fwt}. In this article, we demonstrate that 5D PBH evaporation
into such sterile state provides an economic way to explain the
KM3-230213A neutrino event without conflicting with gamma-ray
observation limits.

The layout of the paper is as follows. In Sec.~\ref{sec:2} we summarize the
generalities of serile-active neutrino oscillations in scenarios with
large extra dimensions. Along the way, we also briefly discuss the
possibility of sterile neutrinos taking shortcuts in extra dimensions, which leads to distinct oscillation signatures~\cite{Pas:2005rb}. In Sec~\ref{sec:3} we describe our model in
detail and define its parameter space. After that we take benchmark
points of this space to demonstrate (provide proof of concept) that 5D PBH evaporation can
explain the origin of KM3-230213A. The paper wraps up with some conclusions presented in Sec.~\ref{sec:4}.

\section{Sterile-active oscillation with shortcut through extra dimension}
\label{sec:2}

A light sterile neutrino with a mass $\sim 1~{\rm eV}$ continues to be
interesting because of multiple clues from terrestrial
experiments~\cite{Diaz:2019fwt}. This hypothesis, however, suffers from strong constraints
from IceCube data~\cite{IceCubeCollaboration:2024dxk} and is in tension with cosmological
observations~\cite{Hagstotz:2020ukm}. Consistency with the most restrictive bounds from
IceCube data requires $\Delta m_{14}^2 \sim 1~{\rm eV}^2$ and
$\sin^2 (2 \theta_{24}) \sim 10^{-2}$, where
$\Delta m^2_{ij} = m^2_i - m^2_j$ describes the mass difference between
neutrino mass eigenstates $i,j$ and $\theta_{ij}$ defines how much
these mass eigenstates mix to form neutrino flavors, with $i=1,2,3,4$~\cite{IceCubeCollaboration:2024dxk}.

Beyond this observational dilemma, the mechanism behind neutrino
masses remains unknown. A compelling explanation emerges from models
featuring flat large extra dimensions~\cite{Dienes:1998sb, Arkani-Hamed:1998wuz,
  Dvali:1999cn}. Within these models
 right-handed neutrinos are allowed to propagate in the
 higher-dimensional bulk. The large volume of the extra dimensions
 causes a suppression of the Yukawa couplings, which naturally
 produces light Dirac neutrinos. A key by-product of this framework is
 the appearance of a KK neutrino tower. The masses of these KK states
 are inversely proportional to the compactification radius. In the dark dimension scenario, with $R_\perp$ around a micrometer, the KK states could mix with the SM neutrinos~\cite{Machado:2011jt,Forero:2022skg}. Actually,
 when it comes to mixing with active SM neutrinos, it is only the
 lower mass states of the tower that play a relevant role. Besides, the 
 addition of bulk masses suppresses the mixing of the first KK modes
 to active neutrinos allowing such models to remain consistent with
 experimental data~\cite{Carena:2017qhd,Anchordoqui:2023wkm}.
 
Motivated by the preceding discussion, herein we adopt a minimalistic
approach and consider the extension of the SM by one sterile
neutrino propagating into the bulk. For the considerations in the present work, the active-sterile
oscillation can be safely approximated by a two-state system. We take as benchmark a mass-squared splitting $\delta m^2 \sim 1~{\rm
  eV}$ and we set the mixing angle to $\theta \sim 10^{-2}$.

\begin{figure}[htb!]
   \postscript{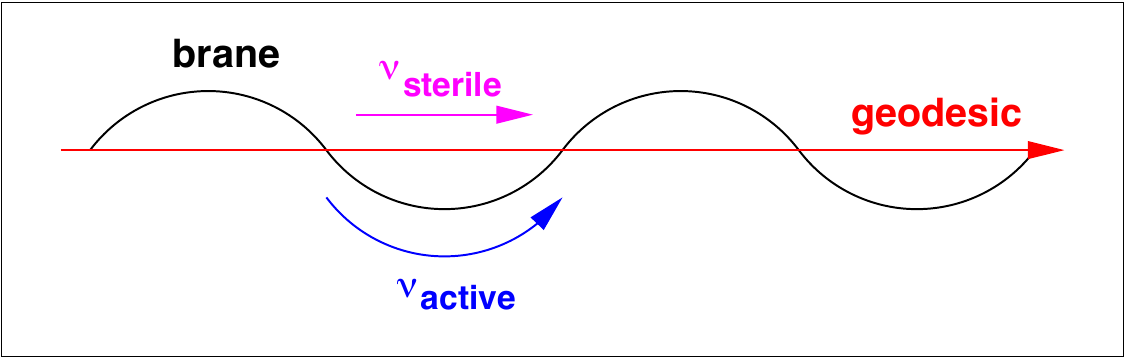}{0.9}
   \caption{Schematic representation of the propagation of a sterile
     neutrino through a geodesic in the bulk and an active neutrino confined to the brane~\cite{Pas:2005rb}.
   \label{fig:0}}
\end{figure}

Two possibilities emerge in the extra-dimensional brane-world
scenario~\cite{Pas:2005rb}. On the one hand, if the brane is rigid and flat in its embedding, then the
geodesic of the sterile neutrino is the same as the geodesic of
the active neutrino on the brane. On the other hand, if the brane is allowed to
fluctuate in its embedding, then the sterile neutrino may have a
different trajectory, with a shorter geodesic than that of the active
neutrino constrained to the brane, see Fig.~\ref{fig:0}. This implies that if the active neutrino propagates on a non-rigid brane
there are two sources of phase difference causing oscillation: {\it (i)}~the usual kinematic one
coming from the neutrino mass-squared difference, and {\it (ii)}~a
geometric one coming from the difference in travel times of the
sterile neutrino through the bulk relative to the active neutrino
confined to the brane. The two phase differences may beat against one
another to produce resonant phenomena.

For neutrinos of energy $E_\nu$, the effective two-flavor mixing angle is given by 
\begin{equation}
 \sin^2(2\tilde{\theta})=\frac{\sin^2(2\theta)}{\sin^2(2\theta)+\cos^2(2\theta)\left[1-\left(E_\nu/E_{\rm
             res} \right)^2 \right]^2} \,,
\end{equation}
where 
\begin{equation}
 E_{\rm res}=\sqrt{\frac{\delta m^2 \ \cos(2\theta)}{2\epsilon}},
\end{equation}
denotes the resonance energy, 
$\delta m^2$ is the mass square difference between the sterile and
active neutrino states, $\theta$ is the usual active-sterile mixing
angle (for a rigid brane) and $\epsilon=\delta t/t$ is the shortcut
parameter, defined as the normalized difference of propagation times
on the brane and in the bulk~\cite{Pas:2005rb}. Since the value of $\epsilon$ is
unknown, the resonance energy could have almost any
value, {\it a priori}. If the KM3-230213A event was amplified through
a resonant enhancement, then
$\epsilon \sim 5 \times 10^{-35}$. In our calculations we adopt this
value. This parameter value have to be eventually explained in a
theory of brane dynamics. Such an explanation exceeds the limitations
of our phenomenological analysis and is reserved for future work.

\begin{figure}[htb!]
   \postscript{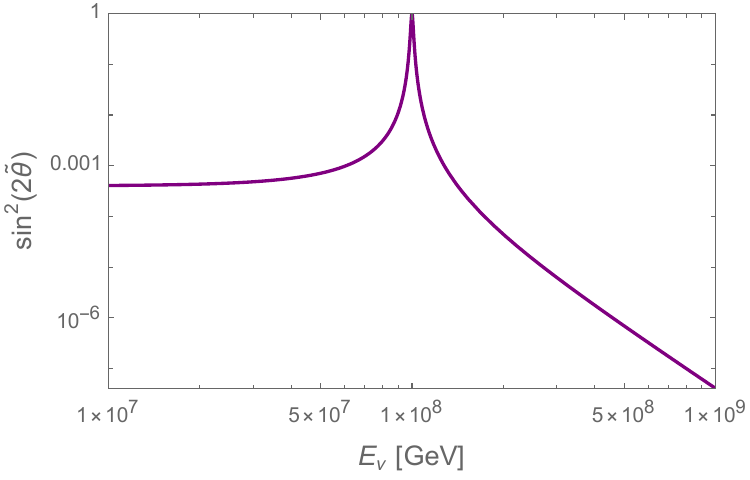}{0.9}
   \caption{Oscillation amplitude $\sin^2 (2\tilde{\theta})$  as a
     function of the neutrino energy $E_\nu$, for a resonance energy
     of $E_{\rm res} = 10^8~{\rm GeV}$. 
   \label{fig:0+}}
\end{figure}

In Fig.~\ref{fig:0+} we show $\sin^2 (2 \tilde \theta)$ for
$\theta \sim 10^{-2}$  as a function of energy. For energies much
larger than the resonance energy $E_{\rm res}$, the
sterile state decouples from the active state, as $\sin^2 (2 \tilde
\theta) \to 0$. For energies much smaller than $E_{\rm res}$, the effective
active-sterile mixing angle is suppressed unless a significant mixing between the active and sterile states occurs. For $0.95 \alt E_\nu/(10^8~{\rm GeV}) \alt 1.05$, we have $\sin^2 (2
\tilde{\theta}) \agt 0.04$.

Putting all the pieces together, the probability of sterile-active
oscillation after a propagation distance $D$ is given by
\begin{equation}
  P_{sa} = \sin^2 (2 \tilde \theta) \ \sin^2 (\delta H \ D/2) \,,
\end{equation}
with
\begin{equation}
  \delta H = \frac{\delta m^2}{2 E_\nu} \sqrt{\sin^2 (2 \theta) +
    \cos^2(2 \theta) \left[1 - \left(\frac{E_\nu}{E_{\rm
            res}}\right)^2\right]^2} \, .
\end{equation}  
For $E_\nu \sim 10^8~{\rm GeV}$, we have
$\delta H \sim 8 \times 10^{-16}~{\rm cm}^{-1}$. Therefore, for a travelling distance $D$ in
the parsec scale, the term $\sin^2(\delta H \, D/2)$ oscillates fast, and
phase-averaging then sets $\langle \sin (\delta H \, D/2)\rangle$ to zero and
$\langle \sin^2 (\delta H \, D) \rangle$ to 1/2. All in all, for the
energy range of interest $0.95
\alt E_\nu/(10^8~{\rm GeV}) \alt 1.05$, the oscillation probability
between sterile and active states can be bounded from below: $P_{sa} \agt 10^{-2}$.

\section{5D PBH evaporation}
\label{sec:3}

Now, in light with our stated plan, we  examine the evaporation of 5D PBHs within the context of the dark dimension. The PBHs perceiving the dark dimension are bigger, colder, and longer-lived than a usual 4D black
hole of the same mass~\cite{Anchordoqui:2022txe,Anchordoqui:2022tgp,Anchordoqui:2024akj,Anchordoqui:2024dxu}. Indeed, the Schwarzschild radius of a $d$-dimensional black
hole scales as~\cite{Tangherlini:1963bw}
\begin{equation}
r_s \sim \frac{1}{M_*} \left(\frac{M}{M_*}\right)^{1/(d-3)} \, ,
\end{equation}  
whereas its temperature scales as
\begin{equation}
  T  \sim M_* \left(\frac{M_*}{M}\right)^{1/(d-3)
  } \, .
  \label{temperatura}
\end{equation}
The black hole lifetime is estimated to be
\begin{equation}
  \tau_H \sim S \ r_s =  \frac{1}{M_*} \
  \left(\frac{M}{M_*}\right)^{(d-1)/(d-3)}  \,,
  \label{tauH5}
\end{equation}
because its entropy scales as~\cite{Anchordoqui:2001cg}
\begin{equation}
S \sim (M/M_*)^{(d-2)/(d-3)} \, . 
\label{entropia}
\end{equation}

\begin{figure}[htb!]
   \postscript{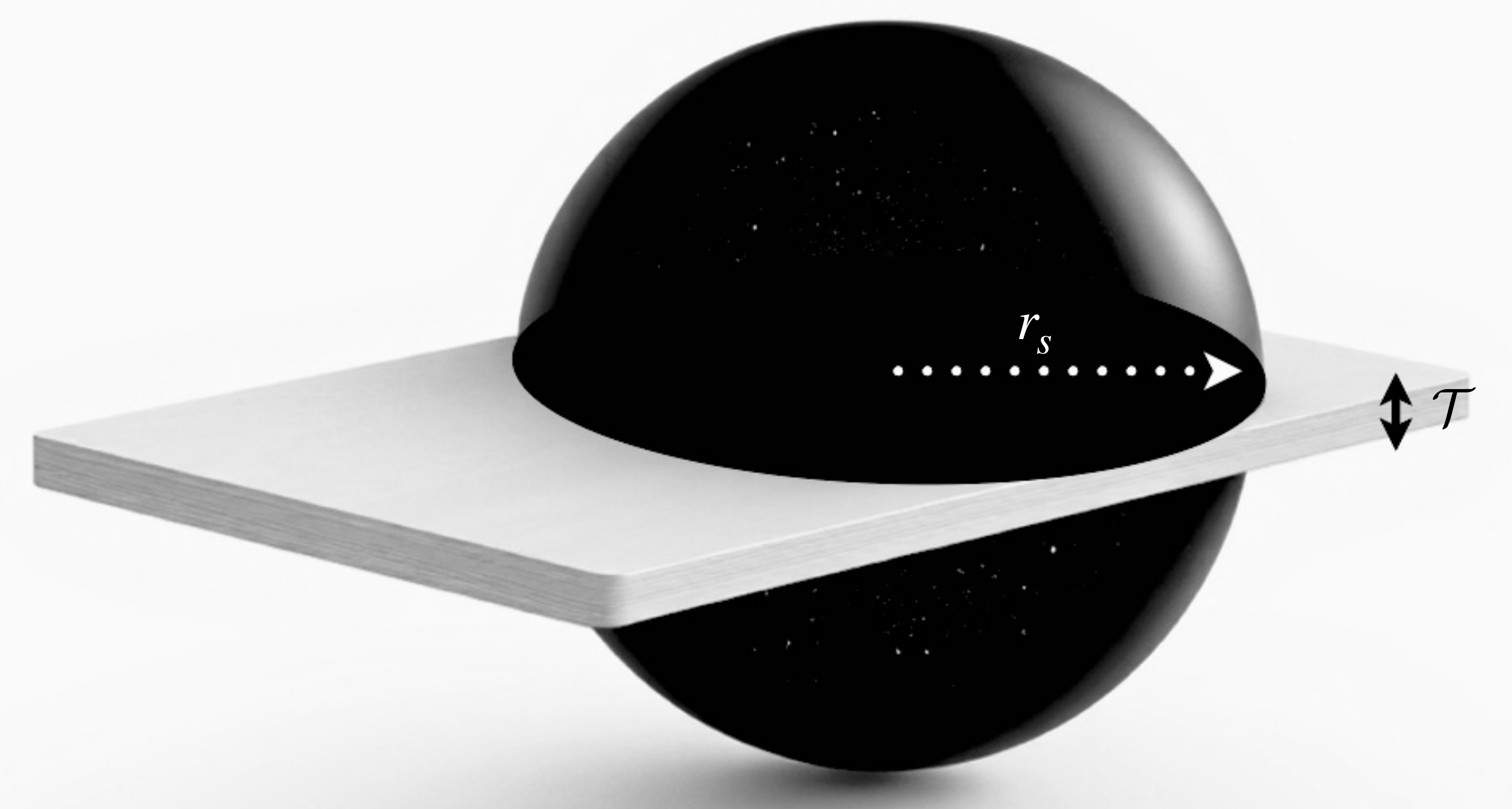}{0.9}
   \caption{Schematic representation of a 5D black hole localized on
    the brane.
   \label{fig:1}}
\end{figure}

Of particular interest here is the relative suppression of black
evaporation into the brane compared to evaporation in the bulk.  The in-brane to bulk phase space suppression
factor can be approximated by $({\cal T}/r_s)^{d-4}$, where  ${\cal T}
\sim 1/M_*$ is the 3-brane thickness that is relevant for SM
excitations~\cite{Argyres:1998qn}. One can check by inspection of
Fig.~\ref{fig:1} that there should be a large suppression
of the SM modes until the 
final-stage emission of the black hole when $r_s$ approaches ${\cal T}$.  Whether the PBH would emit or not
into the brane also depends on whether the PBH stays attached to the
brane as $r_s \to {\cal T}$ or if it can wander off into the
transverse bulk of the dark dimension. Without model dependent assumptions of the bulk and
brane theory it is not possible to calculate the probability of such
wandering in detail. However, as $r_s$ approaches ${\cal T}$, the
recoil effect due to graviton emission may impart the
black hole a relative kick velocity with respect to the brane that would tend to make it escape into the bulk in the final stages of its
life~\cite{Flachi:2005hi,Flachi:2006hw}.
Actually, it is always possible that the PBH formed in the bulk to
start with~\cite{Schwarz:2024tet}. If this were the case, then the PBH would radiate in the bulk (off the brane), where it
cannot emit any of the SM fields localized to the brane. In our calculations we implicitly
assume that 5D PBHs are localized within the dark dimension, away from any branes.

At this stage, to develop some sense for the orders of magnitude
involved, we make contact with experiment. The required number density
of neutrinos for detection of one event per
every hundred years in a km$^2$-area detector can be approximated by
\begin{equation}
n_\nu \sim \frac{1}{A_{\rm eff} \ t} \sim 10^{-30}~{\rm cm^{-3}} \,,
\end{equation}
where $A_{\rm eff} \sim 1~{\rm km}^2$ is the KM3Net effective area~\cite{KM3NeT:2025npi} and where we have increased the
KM3Net time of data taking $t$ by a factor of 100 to accommodate the
tension with IceCube observations ({\it viz}, IceCube has been taking data by
about 10 years with a factor of 10 larger effective area).

Now, using (\ref{temperatura}) it is straightforward to see that a 5D
PBH near the end of its evaporation lifetime and today has a mass $M_\bullet \sim 10^{11}~{\rm
  GeV}$ would emit at a temperature of $10^8~{\rm GeV}$. Furthermore,
using (\ref{entropia}) it follows that such a PBH will almost
instantaneously emit about $S \sim 10^3$ secondaries. The PBH emission
is
thermal in character, following a Planckian distribution with temperature $T_\bullet$. The fraction of
particles emitted in the energy interval ($\Delta E$) of interest $0.95
\alt E /(10^8~{\rm GeV}) \alt 1.05$
is given by
\begin{equation}
  f_{\Delta E} = \frac{\int_{0.95}^{1.05} \frac{x^2}{e^x -1} \ dx}{\int_0^\infty
    \frac{x^2}{e^x -1} \ dx} \sim 10^{-2} \,,
\end{equation}
where $x = E/T_\bullet$.

We assume that the bulk PBH emission consist of: 5 degrees of
freedom of 5D dimensional graviton, 6 degrees of freedom of the
right-handed neutrinos (and corresponding anti-neutrinos), and 4
degrees of freedom of the sterile Dirac fermion. Thus, 27\% of
the particles emitted by the black hole would be eV-scale sterile neutrinos.
The total number of such sterile states emitted per PBH in the range  $0.95
\alt E_\nu /(10^8~{\rm GeV}) \alt 1.05$ is given by
\begin{equation}
  \zeta_{\nu_s} \sim 0.27 \  S  \ f_{\Delta E}
\end{equation}
After oscillation en route to Earth the number of active neutrinos is
\begin{equation}
  \zeta_{\nu_a} \sim  0.27 \  S  \ f_{\Delta E} \ P_{sa} \sim
  0.027 \, .
\end{equation}  
All told, the number density of these PBHs scales as    
\begin{equation}
  n_{\rm PBH} \sim \frac{1}{\zeta_{\nu_a}} \ \frac{E_\nu}{M_\bullet} \
  n_\nu \sim 4 \times  10^{-32}~{\rm cm}^{-3} \, .
\end{equation}
Thus, the required energy density in PBHs to accommodate the KM3-230213A
neutrino is estimated to be
\begin{equation}
\rho_{\rm PBH} \sim M_\bullet \ n_{\rm PBH} \sim 
4 \times 10^{-21}~{\rm GeV/cm}^3 \, .
\label{rhopbh}
\end{equation}
By comparing (\ref{rhopbh}) with the local dark matter
density
\begin{equation}
  \rho_{\rm DM} \sim 0.4~{\rm GeV/cm}^3 \, ,
\end{equation}
we estimate the fraction of PBH emitting neutrinos relative to the
total dark matter density to be pocket-size:
\begin{equation}
  f_{\rm PBH} \equiv \frac{\rho_{\rm PBH}}{\rho_{\rm DM}} \sim
  10^{-20}  \, .
\label{21}
\end{equation}

At this point it is pertinent to
connect with
IceCube searches of superheavy dark
matter particles $X$ decaying to neutrinos. The
IceCube Collaboration reported lower limits on the lifetime of the $X$ particles on the assumption that $X$'s are clustered in the
Galactic halo and that the superheavy dark matter density saturates
the dark matter density~\cite{IceCube:2018tkk}. These null results can
then be translated into a bound of the $X$ partial decay width to neutrinos
times the fraction of the superheavy dark matter relative to
the total dark matter density~\cite{Guepin:2021ljb}. For PBHs, the
IceCube bound can be reinterpreted as
\begin{equation}
  \Gamma_{{\rm PBH}} \ f_{\rm PBH} \alt
  10^{-29.7}~{\rm  s}^{-1} \,,
\label{22}
\end{equation}  
where
\begin{equation}
  \Gamma_{{\rm PBH}} \sim 10^{-17}  \left(\frac{M}{10^{10.4}~{\rm g}}
    \right)^{-2}~{\rm s}^{-1}
\label{23}
  \end{equation}    
is the PBH decay
width. To derive (\ref{23}) we have used (\ref{tauH5}), taking the age of
the universe to be ${\cal O} (10^{17}~{\rm s})$. When (\ref{21}) and (\ref{23}) are substituted into (\ref{22}), the IceCube bound is straightforwardly satisfied. 

At this stage, it is constructive to connect with contrasting and
complementary perspectives to comment on caveats of our analysis. The expected signal from Hawking emission is broad in
energy, producing a wide primary spectrum of gravitons and sterile
neutrinos. Nevertheless, as shown in Sec.~\ref{sec:2}, a line-like feature
would emerge via the resonant effect of neutrino oscillation if
sterile states take shortcuts through the extra dimension. The
suppression of the sterile-active oscillation at lower energies
naturally depends $\theta$. A slight adjustment to the mixing angle $\theta$
may be required to align with the observed data. However, we emphasize that a
substantial region within the allowed parameter space could provide
considerable flexibility for this adjustment, given that $f_{\rm PBH}
\sim 10^{-20}$ and $\Gamma_{\rm BH} \, f_{\rm BH} \sim 10^{-37}$ is more
than eight orders of magnitude below the IceCube bound, where we have
taken $M \sim 10^{10.4}~{\rm g}$.

Besides, the most immediate impact of the discovery
of astrophysical neutrinos is that the flux level observed is exceptionally high by astronomical
standards. As a matter of fact, the magnitude of the flux measured with IceCube may be as large as
  $10^{-7}~{\rm GeV \ cm^{-2} \ s^{-1} \ sr^{-1}}$ around
  30~TeV~\cite{IceCube:2015gsk}. This finding leads directly to the inference that if sources of TeV-PeV
  neutrinos are transparent to $\gamma$-rays, a strong tension with the
  isotropic diffuse $\gamma$-ray background measured by the {\it
    Fermi} satellite~\cite{Fermi-LAT:2014ryh} becomes unavoidable -- independently of the
  neutrino production mechanism~\cite{Murase:2015xka}. Therefore, it
  is evident that the sources of the
  astrophysical neutrinos observed by IceCube must be hidden cosmic ray
  accelerators, in which neutrinos are produced through $\pi^\pm$ decay, but the photon emission from $\pi^0$ decay is
  absorbed by interactions at the
  source~\cite{Murase:2015xka,Fang:2022trf}. As a sharp reader might
  have noticed, the emitted radiation from evaporation of bulk PBHs
  resembles that of astrophysical hidden sources. A reliable estimate
  of the best fit paramaters that can accommodate the
high-energy ($\agt 10~{\rm TeV}$) spectra of photons and neutrinos
together with their corresponding upper limits demands a sophisticated
numerical analysis. Such an ambitious project, however, is beyond the scope of this paper.
  
\section{Conclusions}
\label{sec:4}

We have proposed a new model to explain the origin of the
KM3-230213A neutrino~\cite{KM3NeT:2025npi}. Our model is based on the evaporation of 5D PBHs
within the dark dimension scenario. We have assumed that these black
holes live in the bulk and thus only
radiate gravitons and sterile neutrinos. Moreover, in the last stage
of evaporation they radiate particles with energies just above the Hagedorn temperature. Since the species scale in the dark dimension
scenario is ${\cal O}(10^{9}~{\rm GeV})$ the average energy of the radiated
sterile neutrinos is naturally consistent with the energy of
KM3-230213A. These sterile states could  take shortcuts through the dark
dimension and oscillate en route to Earth into
active neutrinos. Our proposal is also consistent with IceCube upper limits on the
partial decay width of superheavy dark matter particles into neutrinos~\cite{IceCube:2018tkk}. 

A point worth noting at this juncture is the role played by the dark
dimension in our proposal. One may wonder why we did not adopt a more
economic perspective and argue that the photon-to-neutrino ratio of 4D PBH
emission can be suppressed via beyond SM physics. For example, to
explain neutrino oscillations a minimal extension of the SM is to
introduce three right-handed neutrinos along with their Yukawa
interactions with the Higgs field to generate Dirac neutrino
masses. In principle the right-handed neutrinos can live in four
dimensions. This minimal SM extension leads to $\gamma \div \nu =
1\div6$. This ratio can be further suppressed in 4D supersymmetric models~\cite{Halzen:1991uw}, where
e.g. light modulinos can oscillate into SM
neutrinos~\cite{Benakli:1997iu} and of course with the addition of
sterile neutrinos. Though such a 4D extension would suppress direct
photon emission, hadronization of quarks and gluons would still produce a
flux of $\pi^0$ which on decay could give rise to a photon signal. Markedly different,
 5D PBHs radiate gravitons and sterile neutrinos in the bulk for most
 of their life-time, and only in the last stage of evaporation they radiate SM modes with energies just above the Hagedorn temperature. Thus, the way in which
5D PBHs localized on the brane evolve over time could help to reduce (though not fully
eliminate) the tension with non-observation of a photon signal.

Indeed, using Eq.~(\ref{4DT}) it is straightforward to see that 4D PBHs of $M \sim 10^{39}~{\rm GeV}$  Hawking radiate at
$T \sim 5~{\rm MeV}$. As a consequence, such 4D PBHs
would emit ${\cal O}(10^{39}~{\rm GeV}$) of energy into dangerous energetic modes. In sharp contrast, even if 5D  PBHs stay localized on
the brane, they emit at most ${\cal O}(10^{11}~{\rm GeV})$ amount of energy
into visible SM modes, a suppression factor of $10^{-28}$ relative to
4D PBHs of the same mass. Correspondingly, all 
evaporation constraints on the density of
PBHs~\cite{Carr:2020xqk,Green:2020jor,Villanueva-Domingo:2021spv} are significantly weakened
in the dark dimension scenario. As might be expected, however, if 5D PBHs live
in the bulk then there is no emission into visible SM modes on the
brane. Thus, we can safely conclude that
evaporation of bulk PBHs can explain the origin of the KM3-230213A neutrino, while evading constraints from upper
limits on the gamma-ray flux. The model predicts a distinctive 
neutrino-line feature without a photon counterpart. Future observations by the IceCube neutrino telescope, as well as
those from next generation neutrino experiments, may provide a
definitive answer to the ideas introduced in this paper.

In closing, we note that the relative suppression of black hole radiation into the brane
compared to gravitons radiated off into the bulk would also
characterized 5D memory-burdened PBHs. As a consequence, evaporation
constraints on the density of memory-burdened
PBHs~\cite{Alexandre:2024nuo,Thoss:2024hsr,Zantedeschi:2024ram,Montefalcone:2025akm,Chianese:2025wrk}
are also significantly weakened
within the dark dimension scenario. In particular, constraints from
$\gamma$-ray experiments and primordial nucleosynthesis light element
destruction~\cite{Thoss:2024hsr} are alleviated completely even for $k = 1$ (i.e., a
slowdown of evaporation by only a single power of
entropy). Memory-burdened PBHs perceiving the dark dimension could thus
constitute all of the dark matter in the universe if $M \agt 100~{\rm
  g}$ and perhaps be the source of the KM3-230213A neutrino.

\section*{Acknowledgments}

We thank Alek Bedroya and David Kaiser for  discussions. The research of L.A.A. is supported by the U.S. National Science
Foundation (NSF) grant PHY-2412679. The research of F.H. is supported
in part by the NSF under grants PHY-2209445 and OPP-2042807. The work of D.L. is
supported by the Origins Excellence Cluster and by the
German-Israel-Project (DIP) on Holography and the Swampland.

\end{document}